\documentclass[reprint,twocolumn,amsmath,amssymb,aps,nofootinbib,pre]{revtex4-2}

\usepackage[dvipdfmx]{graphicx,hyperref}
\usepackage{dcolumn}
\usepackage{bm}
\usepackage{xcolor}
\usepackage{ulem}
\usepackage{mathtools}
\usepackage{comment}
\allowdisplaybreaks
\newcommand{\pprime}{{\prime\prime}}

\begin{document}
\title{Kinetic theory of dilute weakly charged granular gases\\
with hard-core and inverse power-law interactions under uniform shear flow}
\author{Yuria Kobayashi}
\affiliation{
    Department of Mechanical Systems Engineering, 
    Tokyo University of Agriculture and Technology,
    2-24-16, Naka-cho, Koganei, Tokyo 184-8588, Japan}
\author{Makoto R. Kikuchi}
\affiliation{
    Department of Mechanical Systems Engineering, 
    Tokyo University of Agriculture and Technology,
    2-24-16, Naka-cho, Koganei, Tokyo 184-8588, Japan}
\author{Shunsuke Iizuka}
\affiliation{
    Department of Mechanical Systems Engineering, 
    Tokyo University of Agriculture and Technology,
    2-24-16, Naka-cho, Koganei, Tokyo 184-8588, Japan}
\author{Satoshi Takada}
\email[Corresponding author, e-mail: ]{takada@go.tuat.ac.jp}
\affiliation{
    Department of Mechanical Systems Engineering, 
    Tokyo University of Agriculture and Technology,
    2-24-16, Naka-cho, Koganei, Tokyo 184-8588, Japan}
\date{\today}
\begin{abstract}
We develop a kinetic-theory framework to investigate the steady rheology of a dilute gas interacting via a repulsive potential under uniform shear flow.
Starting from the Boltzmann equation with a restitution coefficient that depends on the impact velocity and potential strength, we derive evolution equations for the stress tensor based on Grad's moment expansion.
The resulting expressions for the collisional rates and transport coefficients are fitted with simple analytical functions that capture their temperature dependence over a wide range of shear rates.
Comparison with direct simulation Monte Carlo (DSMC) results shows excellent quantitative agreement for the shear stress, temperature anisotropy, and shear viscosity.
We also analyze the velocity distribution functions, revealing that the system remains nearly Maxwellian even under strong shear.
\end{abstract}
\maketitle

\section{Introduction}
Charged granular gases, in which macroscopic particles carry net electrical charges and interact through long-range Coulombic or screened-Coulomb forces in addition to direct inelastic collisions, represent a unique class of out-of-equilibrium many-body systems. 
Such systems bridge granular matter and plasma or colloidal physics~\cite{Ivlev12}, where the dissipative dynamics characteristic of granular flows coexist with energy and momentum transfer mediated by electrostatic interactions. 
They are of broad relevance, ranging from triboelectric powder handling and electrostatic separation to natural phenomena such as volcanic plumes and atmospheric dust electrification~\cite{Fortov04, Lacks11, Cimarelli14}.

For neutral granular gases, kinetic theory based on the (inelastic) Boltzmann or Enskog equation has provided a firm theoretical foundation \cite{Jenkins85, Brey98, Garzo99, Goldhirsch03}. 
In particular, the homogeneous cooling state (HCS) and uniform shear flow (USF) have been extensively analyzed, leading to quantitative predictions of transport coefficients, non-Newtonian viscosity, and normal stress differences~\cite{Brey98, Garzo99, Saha14}. 
These theories have been validated by both simulations and experiments up to moderate densities and restitution coefficients~\cite{Mitarai07, Chialvo13}.

By contrast, the rheology of charged granular gases remains much less explored. Long-range electrostatic repulsion modifies collision frequencies, relative velocities, and spatial correlations, introducing an additional energy scale and a new dimensionless control parameter-the ratio of electrostatic potential energy to kinetic energy~\cite{Scheffler02, Poschel03, Takada17, Singh18, Singh19, Takada22}. 
In freely cooling charged systems, previous works \cite{Scheffler02, Poschel03, Takada17, Takada22} demonstrated that Coulomb repulsion suppresses collision rates, leading to slower-than-Haff cooling and modified transport coefficients. 
Yet, these studies were limited mainly to homogeneous, unforced systems.

In the present work, we focus on the regime of weakly charged granular particles, where mechanical inelastic contact remains the dominant dissipation mechanism in binary collisions, while the electrostatic interaction acts as a perturbative long-range correction. 
In other words, the typical electrostatic potential energy at contact is assumed to be smaller than or comparable to the granular kinetic energy. 
Strongly charged regimes, in which long-range Coulomb interactions suppress direct mechanical contact or lead to collective plasma-like effects, are beyond the scope of the present kinetic description.

Only a few attempts have addressed the effect of electrostatic interactions under steady shear~\cite{Yoshii23, Takada24, Kobayashi25}. 
Yoshii {\it et al.}~\cite{Yoshii23} analyzed a dilute binary granular mixture interacting through an effective square-shoulder potential, motivated by coarse-grained electrostatic repulsion, and derived shear-rate-dependent viscosity within kinetic theory. 
Their results suggested a crossover from hard-sphere-like (Bagnold) scaling at high shear to deviation regimes at lower shear. 
However, a systematic first-principles kinetic description for a single-component charged granular gas, explicitly incorporating both hard-core and inverse-power-law interactions, has not been developed.

In this study, we extend kinetic theory to formulate the non-Newtonian rheology of a dilute charged granular gas subjected to steady simple shear. 
The remainder of this paper is organized as follows. 
Section~\ref{sec:model} introduces the model and kinetic setup.
Section~\ref{sec:scattering} formulates the scattering process, and then Boltzmann equation is analyzed in Sect.~\ref{sec:Boltzmann}.
Section~\ref{sec:rheology} presents steady-state rheology and comparison with simulations.
And finally, Sect.~\ref{sec:conclusion} concludes with a summary.
Appendix \ref{sec:theta} presents the explicit expression of the collision angle for some cases.

\section{Model and setup}\label{sec:model}
\begin{figure}[htbp]
    \centering
    \includegraphics[width=\linewidth]{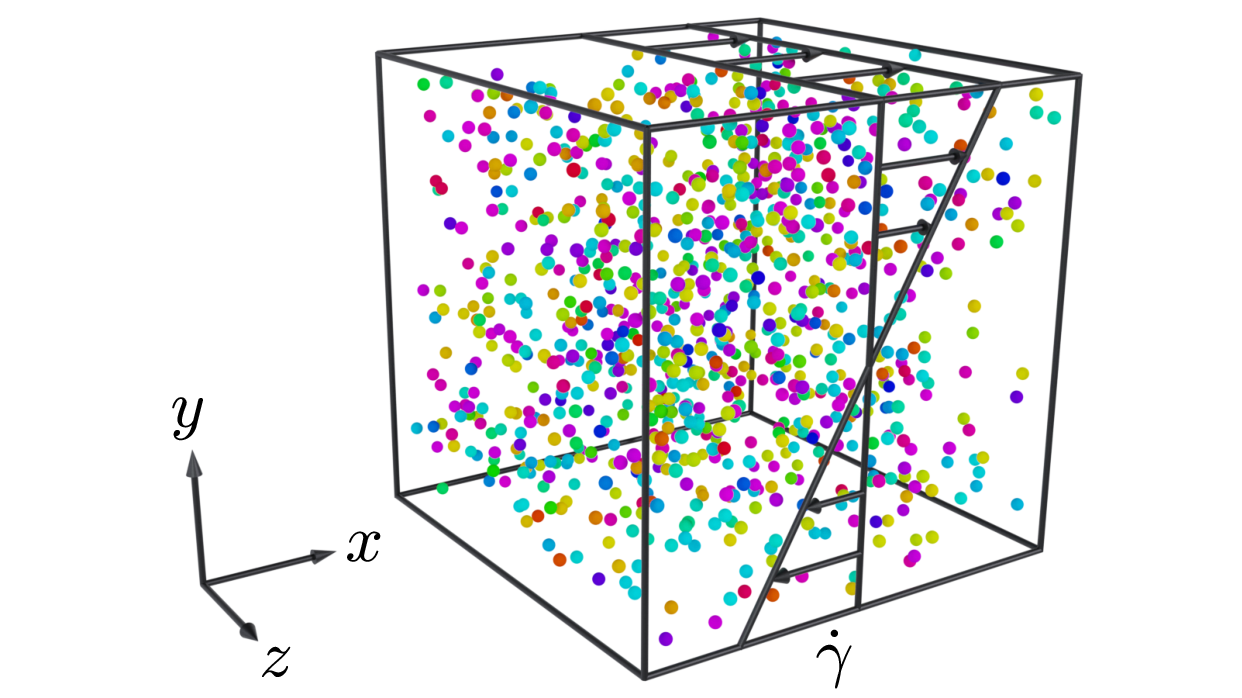}
    \caption{Schematic of our system.
    Monodisperse particles are randomly distributed in a cubic box.
    A shear is applied with the shear rate $\dot\gamma$.}
    \label{fig:setup}
\end{figure}

\begin{figure}[htbp]
    \centering
    \includegraphics[width=0.65\linewidth]{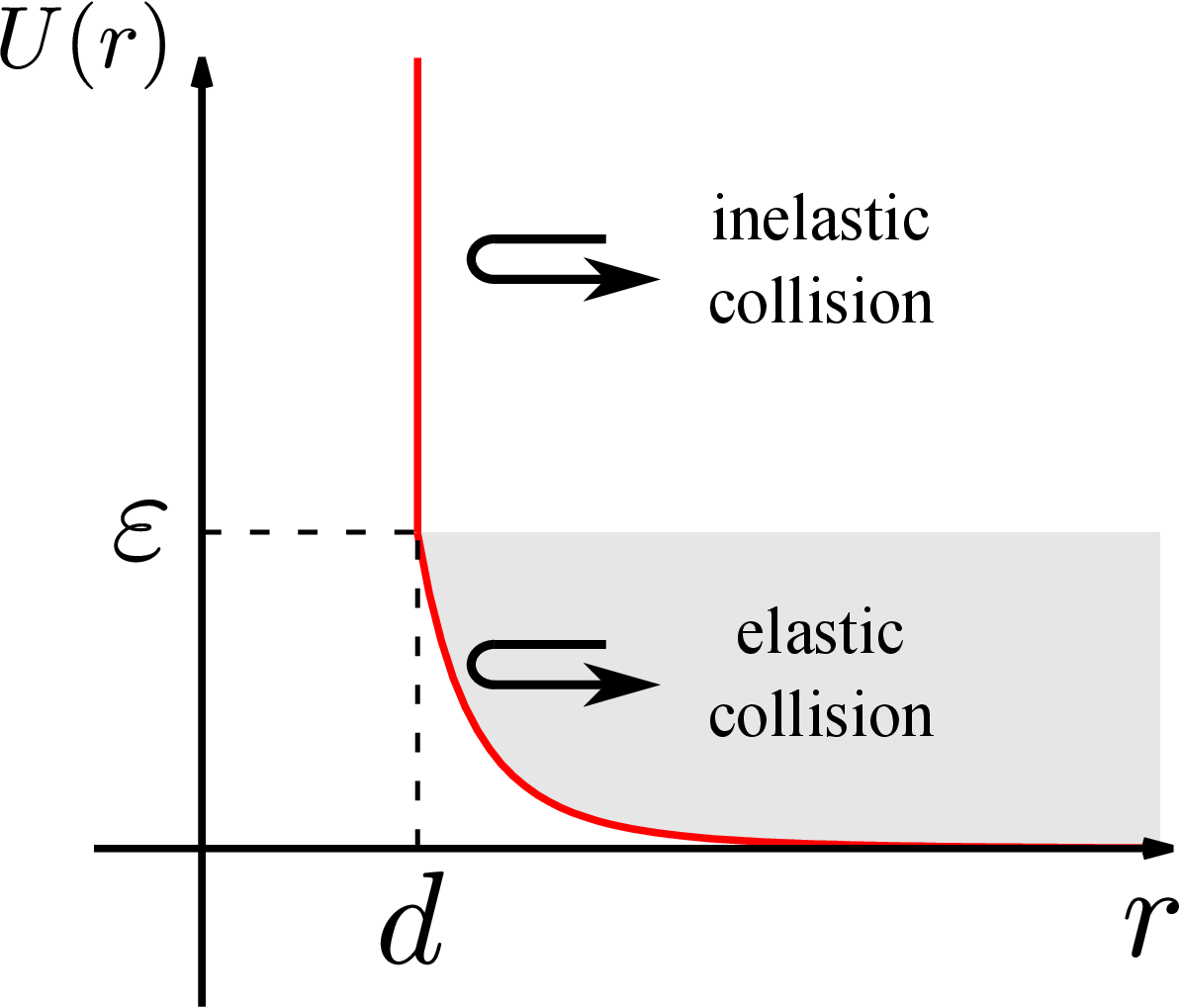}
    \caption{The interparticle potential used in this paper.
    The shaded and non-shaded region represents whether a collision is elastic or inelastic.}
    \label{fig:potential}
\end{figure}

Let us explain the model and setup.
The monodisperse particles whose mass and radius are $m$ and $d$, respectively, are randomly distributed in the cubic box with the linear size $L$ as shown in Fig.~\ref{fig:setup}.
Now, the number density $n$ is set to be sufficiently small ($nd^3\ll 1$).
The interparticle potential $U(r)$ is assumed to be given by the sum of the hard-core and the inverse power-law potentials as
\begin{equation}
    U(r) = 
    \begin{cases}
        \infty & (r\le d)\\
        \varepsilon\left(\dfrac{d}{r}\right)^\alpha & (r>d)
    \end{cases},
    \label{eq:potential}
\end{equation}
(see Fig.~\ref{fig:potential}) where $r$ is the distance between particles, $\varepsilon$ characterizes the magnitude of the repulsion, and $\alpha$ is the exponent of the inverse power-law potential.
Here, we assume $\alpha<2$, which ensures that $4\pi r^2U(r)\to 0$ as $r\to \infty$.
When two particles collide with each other at $r=d$, there occurs an inelastic collision with the restitution coefficient $e\,(<1)$ as shown in Fig.~\ref{fig:potential}.

We now apply a uniform shear with shear rate $\dot{\gamma}$ to the system.
As illustrated in Fig.~\ref{fig:setup}, the shear flow is imposed along the $x$-direction with the velocity gradient in the $y$-direction.
In this case, the deviation of the particle velocity from the uniform shear flow is characterized by
\begin{equation}
    \bm{V}_i 
    \equiv \bm{v}_i - \dot\gamma y_i \hat{\bm{e}}_x, 
    \label{eq:def_V}
\end{equation}
where $\bm{V}_i$ represents the peculiar velocity of particle $i$ relative to the imposed shear.

\section{Scattering process}\label{sec:scattering}
\begin{figure}[htbp]
    \centering
    \includegraphics[width=0.75\linewidth]{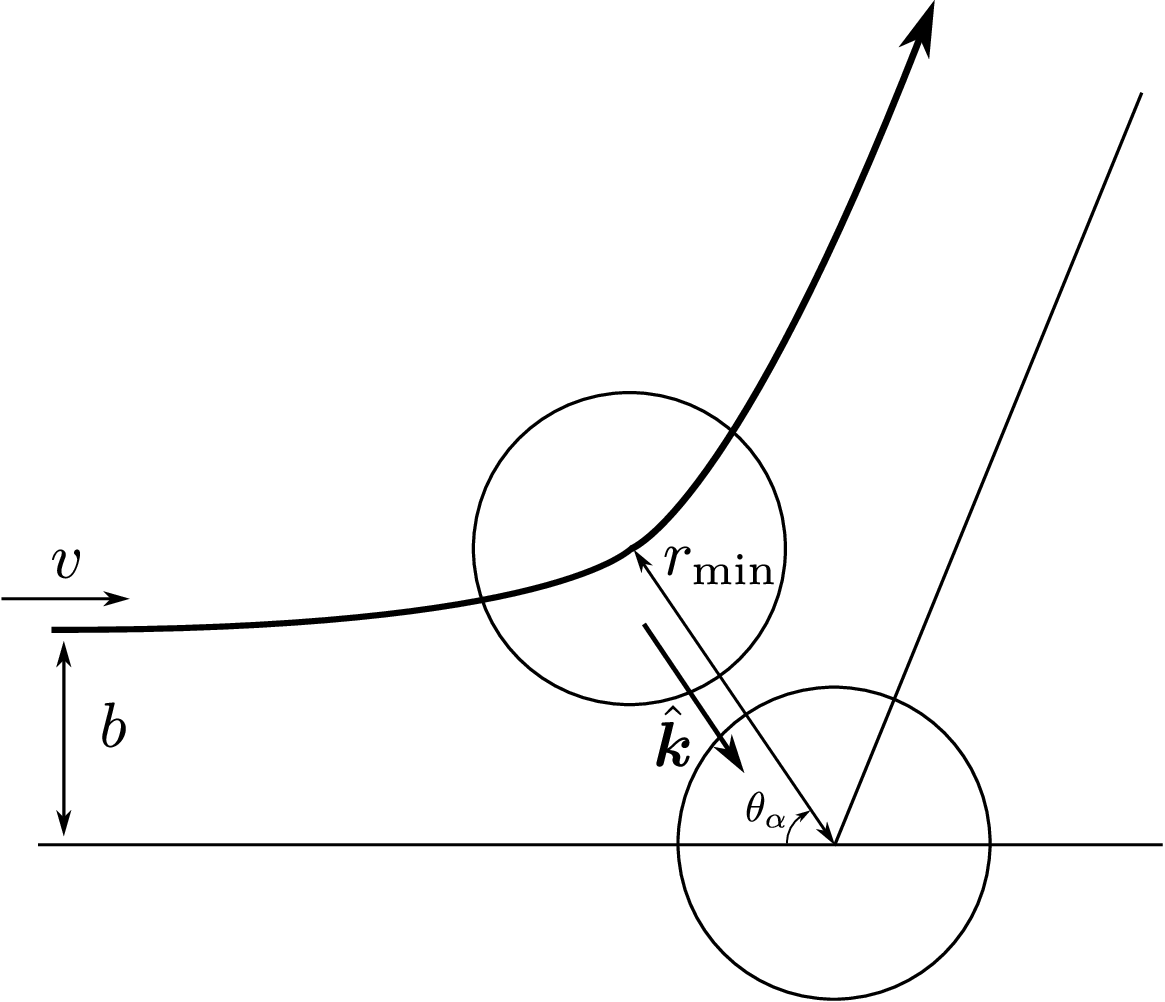}
    \caption{A schematic of a scattering process.
    Two particles are colliding with each other with the impact parameter $b$ and the relative speed $v$.}
    \label{fig:scattering}
\end{figure}
In this section, let us calculate the collision angle by considering the scattering process as shown in Fig.~\ref{fig:scattering}.
Let us consider the situation that two particles approach to each other with the relative speed $v$ and the impact parameter $b$.
The angle $\theta_\alpha$ between the point of closest approach and infinity is given by~\cite{Goldstein}
\begin{align}
    \theta_\alpha
    &= b \int_0^{1/r_\mathrm{min}}
    \frac{\mathrm{d}u}{\sqrt{1-b^2u^2-\dfrac{4}{mv^2}U\left(\dfrac{1}{u}\right)}}\nonumber\\
    &= \int_0^{x_0}
    \frac{\mathrm{d}x}{\sqrt{1-x^2-\left(\dfrac{x}{\tilde{b}}\right)^\alpha}},
    \label{eq:theta}
\end{align}
where we have introduced $x=bu$, $r_\mathrm{min}\ge d$ is the closest distance between the particles, $x_0=b/r_\mathrm{min}$, and 
\begin{equation}
    \tilde{b}\equiv
    \left(\frac{mv^2}{4\varepsilon}\right)^{1/\alpha}\frac{b}{d}.
\end{equation}
We note that the analytical calculation is available for a certain value of $\alpha$~\cite{Goldstein, Takada24, Kobayashi25} (see Appendix \ref{sec:theta}).

Depending on the values of $v$ and $b$, two types of collision can occur.
The first is an inelastic collision with restitution coefficient $e$ at the hard core, and the second is an elastic collision.
From energy conservation, the inelastic collision takes place if
\begin{equation}
    v \ge \sqrt{\dfrac{4\varepsilon}{m}}, \quad
    b \le \nu_\mathrm{r} d,
\end{equation}
where $\nu_\mathrm{r}$ is the refractive index defined as
\begin{equation}
    \nu_\mathrm{r}
    \equiv
    \sqrt{1 - \dfrac{4\varepsilon}{mv^2}}.
\end{equation}
Otherwise, the collision is elastic.

In the case of an inelastic collision, the particles rebound at the hard-core potential with restitution coefficient $e$.
It should be noted, however, that the relative velocity at the moment of collision is reduced compared with the initial relative velocity $v$ at infinity, due to the deceleration induced by the repulsive potential.
On the other hand, in the elastic collision, no dissipation occurs and the restitution coefficient is simply unity.
When the collision is viewed ``macroscopically,'' i.e. in terms of the state sufficiently far from the interaction region, the effective restitution coefficient $\mathcal{E}$ is given by
\begin{equation}
    \mathcal{E} =
    \begin{cases}
    \sqrt{1 - (1 - e^2)\nu_\mathrm{r}^2} & \left(v \ge \sqrt{\dfrac{4\varepsilon}{m}},\ b\le \nu_\mathrm{r} d\right)\\
    1 & \left(\text{otherwise}\right)
    \end{cases},
    \label{eq:macroscopic_e}
\end{equation}
so that $\alpha$ encapsulates both the microscopic dissipation and the macroscopic scattering behavior.
Physically, the weak dependence of the effective restitution coefficient $\mathcal{E}$ on $e$ arises because most collisions are either fully inelastic (high-$T$ limit) or nearly elastic (low-$T$ limit), with the transition region contributing only marginally to the average energy dissipation rate.
This behavior implies that the rheology is dominated by the repulsive potential rather than by microscopic dissipation details.

Following this formulation, one can calculate the dependence of the scattering angle $\theta$ on the relative velocity $v$ and the impact parameter $b$ for each power $\alpha$.
As explained before, $\theta$ can be obtained analytically for specific cases of $\alpha$.
In Appendix \ref{sec:theta}, we present explicit results for $\alpha = 4$ and $\alpha = 6$.

\begin{figure}[htbp]
    \centering
    \includegraphics[width=\linewidth]{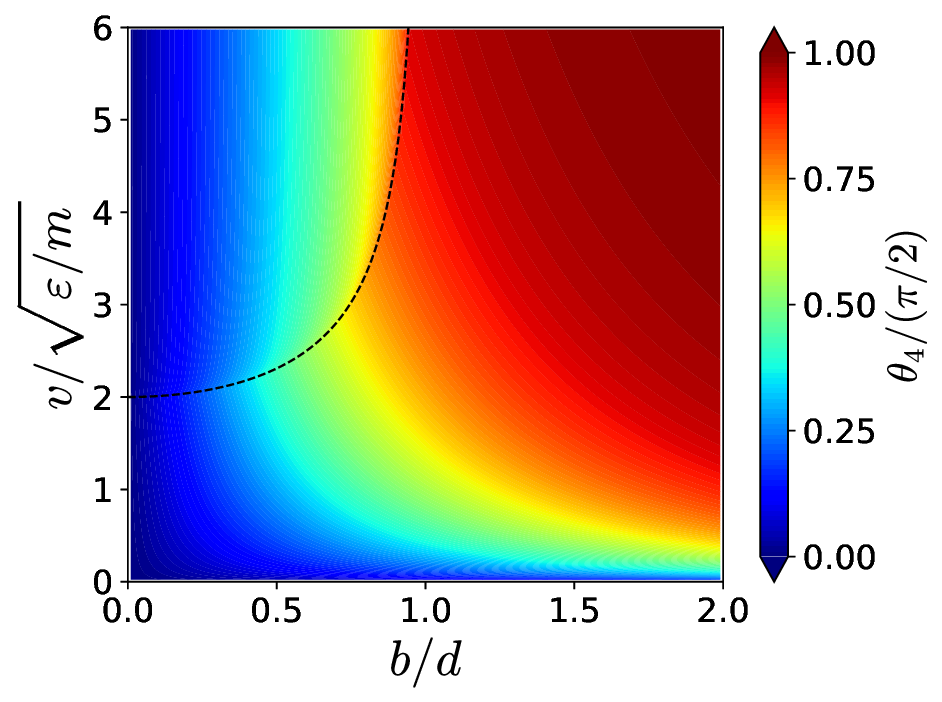}
    \caption{Dependence of the collision angle $\theta_\alpha$ on the relative speed $v$ and impact parameter $b$ for $\alpha=4$. The color scale represents the magnitude of the angle.
    The dashed line represents Eq.~\eqref{eq:b_boundary}.}
    \label{fig:theta}
\end{figure}

Figure \ref{fig:theta} shows the dependence of the scattering angle $\theta_\alpha$ on $v$ and $b$ for $\alpha=4$.
For sufficiently large relative velocities or large impact parameters, the repulsive interaction becomes negligible and the particle trajectories remain nearly straight.
In this regime, the scattering angle approaches $\pi/2$, as expected, which is consistent with the numerical results.
Conversely, for high velocities and small impact parameters, particles collide with the hard core, yielding scattering angles close to those of hard-sphere collisions.
Between these two limiting regimes, the influence of the repulsive potential becomes evident.
Indeed, sharp transitions of the scattering angle are observed in Fig.~\ref{fig:theta}, and the corresponding boundary can be identified as
\begin{equation}
    b = \nu_\mathrm{r} d.\label{eq:b_boundary}
\end{equation}

It should be emphasized that the present model neglects effects of charge redistribution, screening, and multipole interactions, which may become relevant at higher densities or in partially ionized media.
Nevertheless, the model captures the essential physics of Coulomb-suppressed dissipation and provides a tractable kinetic framework for dilute charged granular gases.

\section{Boltzmann equation}\label{sec:Boltzmann}
Using the information of the collision angle $\theta$
introduced in the previous section, we construct the kinetic theory for a dilute gas and derive the time-evolution equation that the stress tensor satisfies.
In a uniform shear system, the particle velocity can be expressed as a deviation from the imposed uniform shear.
The velocity distribution function $f(\bm{v},t)$ is then assumed to obey the Boltzmann equation~\cite{Chapman, Brilliantov, Garzo}
\begin{equation}
    \left(\frac{\partial}{\partial t}
    +\bm{v}_1\cdot \bm\nabla\right)f(\bm{v}_1,t)
    =J(f,f),
    \label{eq:Boltzmann_eq}
\end{equation}
where $J(f,f)$ denotes the collision integral,
\begin{align}
    J(f,f)
    &=\int \mathrm{d}\bm{v}_2 \int d\hat{\bm{k}} \sigma(\chi,\bm{v}_{12}) v_{12}\nonumber\\
    &\hspace{1em}\times
    \left[\frac{1}{\mathcal{E}^2} f(\bm{v}_1^\pprime,t)f(\bm{v}_2^\pprime,t) - f(\bm{v}_1,t)f(\bm{v}_2,t)\right].
\end{align}
Here, the pre-collisional velocity pair $\bm{v}_1^\pprime, \bm{v}_2^\pprime$ and the post-collisional pair $\bm{v}_1, \bm{v}_2$ satisfy the relation
\begin{equation}
    \displaystyle \bm{v}_1 
    = \bm{v}_1^\pprime - \dfrac{1+\mathcal{E}}{2}(\bm{v}_{12}^\pprime \cdot \hat{\bm{k}})\hat{\bm{k}},\quad
    \displaystyle \bm{v}_2 
    = \bm{v}_2^\pprime + \dfrac{1+\mathcal{E}}{2}(\bm{v}_{12}^\pprime \cdot \hat{\bm{k}})\hat{\bm{k}},
    \label{eq:pre_post}
\end{equation}
where $\mathcal{E}$ is the macroscopic restitution coefficient given in Eq.~\eqref{eq:macroscopic_e}.
Here, we do not describe the detailed microscopic dynamics during contact.
Instead, we assume a scattering picture in which the asymptotic pre- and post-collisional states at infinite separation are connected by an effective collision rule characterized by the macroscopic restitution coefficient $\mathcal{E}$.
Under this assumption, the mapping between the pre-collisional velocities $\{\bm{v}_1^\pprime,\bm{v}_2^\pprime\}$ and the post-collisional velocities $\{\bm{v}_1,\bm{v}_2\}$ given by Eq.~\eqref{eq:pre_post} is one-to-one and linear in the normal component of the relative velocity.
Therefore, the Jacobian of the transformation
$\{\bm{v}_1^\pprime,\bm{v}_2^\pprime\}\to
\{\bm{v}_1,\bm{v}_2\}$ can be explicitly evaluated, yielding the factor $1/\mathcal{E}^2$ in the gain term.
As a result, the collision operator is formally equivalent to that of inelastic hard spheres, with the constant restitution coefficient $e$ replaced by the effective coefficient $\mathcal{E}$.

In the case of a uniform shear flow, introducing the peculiar velocity defined in Eq.~\eqref{eq:def_V}, Eq.~\eqref{eq:Boltzmann_eq} can be rewritten as
\begin{equation}
    \left(\frac{\partial}{\partial t}
    -\dot\gamma V_{1,y}\frac{\partial}{\partial V_{1,x}}\right)f(\bm{V}_1,t)
    =J(f,f),
    \label{eq:Boltzmann_eq_shear}
\end{equation}
which corresponds to the standard form of the Boltzmann equation in the uniform-shear frame.
Here, the collision integral implicitly contains information about the repulsive potential through $\mathcal{E}$ and $\theta_\alpha$.
In the dilute limit considered, the stress tensor $P_{k\ell}$ is determined purely by kinetic contributions, allowing direct comparison with kinetic-theory results for uncharged systems.

Multiplying both sides of Eq.~\eqref{eq:Boltzmann_eq} by $mV_{1,k}V_{1,\ell}$ and integrating over $\bm{V}_1$, we obtain~\cite{Santos04, Hayakawa19, Hayakawa17, Takada20, Yoshii23, Kobayashi25}
\begin{equation}
    \frac{\partial}{\partial t}P_{k\ell}
    + \dot\gamma \left(\delta_{k x}P_{y\ell} + \delta_{\ell x}P_{k y}\right)
    = -\Lambda_{k\ell},
    \label{eq:evol_P}
\end{equation}
where
\begin{equation}
    \Lambda_{k\ell}
    \equiv -\int \mathrm{d}\bm{V}_1 mV_{1,k} V_{1,\ell} J(f, f),
\end{equation}
represents the velocity moment of the collision integral, and
\begin{equation}
    P_{k\ell}
    \equiv \int \mathrm{d}\bm{V} 
    m V_k V_\ell f(\bm{V},t),
\end{equation}
is the $k\ell$ component of the stress tensor.
It should be noted that, since we focus only on dilute systems, there is no contribution to the stress from particle contacts.
Using Einstein's summation convention $P_{k}=\sum_{k=x, y, z}^3P_{kk}$ and denoting the number density by $n$, the hydrostatic pressure satisfies $P=P_{kk}/3=nT$, where the temperature $T$ is given by
\begin{equation}
    T= \frac{m}{3n}\int \mathrm{d}\bm{V}\bm{V}^2 f(\bm{V},t).
\end{equation}

From Eq.~\eqref{eq:evol_P}, the quantities of primary interest in dilute systems satisfy the following set of coupled evolution equations~\cite{Santos04, Hayakawa19, Hayakawa17, Takada20}:
\begin{subequations}\label{eq:evol_eqs}
\begin{align}
    \frac{\partial T}{\partial t}
    &= -\frac{2}{3n}\dot\gamma P_{xy} - \frac{\Lambda_{\alpha\alpha}}{3n},\\
    \frac{\partial \Delta T}{\partial t}
    &= -\frac{2}{n}\dot\gamma P_{xy} - \frac{\Lambda_{xx}-\Lambda_{yy}}{n},\\
    \frac{\partial P_{xy}}{\partial t}
    &= -\dot\gamma n\left(T-\frac{1}{3}\Delta T\right) - \Lambda_{xy},
\end{align}
\end{subequations}
where $\Delta T\equiv (P_{xx}-P_{yy})/n$ is the anisotropic temperature.
It should be noted that another anisotropic temperature $\delta T\equiv (P_{xx}-P_{zz})/n$ exists in general situations.
However, $\delta T=\Delta T$ holds in the dilute limit.

Once the set of Eqs.~\eqref{eq:evol_eqs} is solved, the rheological properties of the system can be determined.
However, the explicit form of $\Lambda_{k\ell}$ remains unknown at present.
For this reason, we adopt Grad's approximation for the velocity distribution function~\cite{Grad49, Montanero02, Montanero03, Santos04, Hayakawa19, Hayakawa17, Garzo, Takada20, Yoshii23, Kobayashi25}
\begin{equation}
    f(\bm{V},t)\approx f_\mathrm{M}(\bm{V},t)
    \left[1+\frac{m}{2T}\left(\frac{P_{k\ell}}{nT}-\delta_{k\ell}\right)V_k V_\ell\right],
    \label{eq:Grad}
\end{equation}
with the Maxwellian
\begin{equation}
    f_\mathrm{M}(\bm{V},t)
    = n\left(\frac{m}{2\pi T}\right)^{3/2}
    \exp\left(-\frac{mV^2}{2T}\right),
\end{equation}
which is known to be a good approximation at least for hard-sphere and harmonic potential systems.
Under this assumption, $\Lambda_{k\ell}$ can be expressed explicitly as~\cite{Santos04, Hayakawa19, Hayakawa17, Takada20, Yoshii23, Kobayashi25}
\begin{equation}
    \Lambda_{k\ell}
    = \zeta nT \delta_{k\ell}
    + \nu \left(P_{k\ell}-nT\delta_{k\ell}\right),
\end{equation}
where the coefficients $\zeta$ and $\nu$ are given by
\begin{subequations}
\begin{align}
    \zeta &\equiv \frac{\sqrt{2\pi}}{3}nd^2 v_\mathrm{T} \Omega_{\alpha,5}^{(1)},\\
    \nu &\equiv \frac{\sqrt{2\pi}}{15}nd^2 v_\mathrm{T}
    \left(\Omega_{\alpha,7}^{(1)}+\frac{3}{2}\Omega_{\alpha,7}^{(2)}\right),
\end{align}
\end{subequations}
with the thermal velocity $v_\mathrm{T}\equiv \sqrt{2T/m}$, and we have introduced~\cite{Chapman}
\begin{subequations}\label{eq:Omega_def}
\begin{align}
    \Omega_{\alpha,n}^{(1)}
    &\equiv \int_0^\infty \mathrm{d}g
    \int_0^\infty \mathrm{d}b^*
    \left(1-\mathcal{E}^2\right)b^* g^n
    \cos^2\theta_\alpha \mathrm{e}^{-g^2/2},\\
    \Omega_{\alpha,n}^{(2)}
    &\equiv \int_0^\infty \mathrm{d}g
    \int_0^\infty \mathrm{d}b^*\nonumber\\
    &\hspace{1em}\times
    \left(1+\mathcal{E}\right)^2b^* g^n
    \sin^2\theta_\alpha\cos^2\theta_\alpha
    \mathrm{e}^{-g^2/2}.
\end{align}
\end{subequations}

\begin{figure}[htbp]
    \centering
    \includegraphics[width=\linewidth]{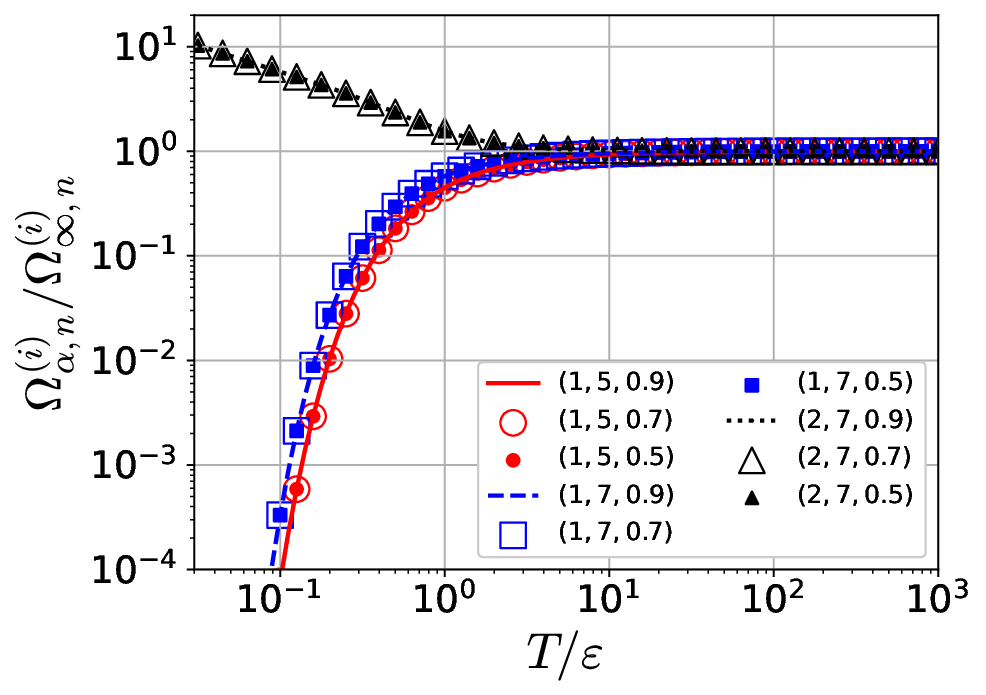}
    \caption{Temperature dependence of $\Omega_{\alpha, n}^{(i)}$ for various sets of $(i, \alpha, e)$ with $\alpha=4$, where $\Omega_{\infty,n}^{(i)}$ represents the hard-core limit of $\Omega_{\alpha, n}^{(i)}$.}
    \label{fig:Omega}
\end{figure}

The temperature dependence of $\Omega_{\alpha,n}^{(1)}$ and $\Omega_{\alpha,n}^{(2)}$ is shown in Fig.~\ref{fig:Omega}.
Interestingly, $\Omega_{\alpha,n}^{(i)}$ exhibits only a weak dependence on the choice of the restitution coefficient $e$.
This can be interpreted as follows:
at high temperatures, almost all collisions are inelastic, whereas at low temperatures they are almost entirely elastic.
In the intermediate regime, a gradual transition occurs.
Since both limiting behaviors coincide, the dependence on $e$ does not become pronounced even in this crossover region.
Based on these results, in the following we focus on the case of restitution coefficient $e=0.9$ and examine the corresponding rheological behavior.

\begin{figure*}[htbp]
    \centering
    \includegraphics[width=\linewidth]{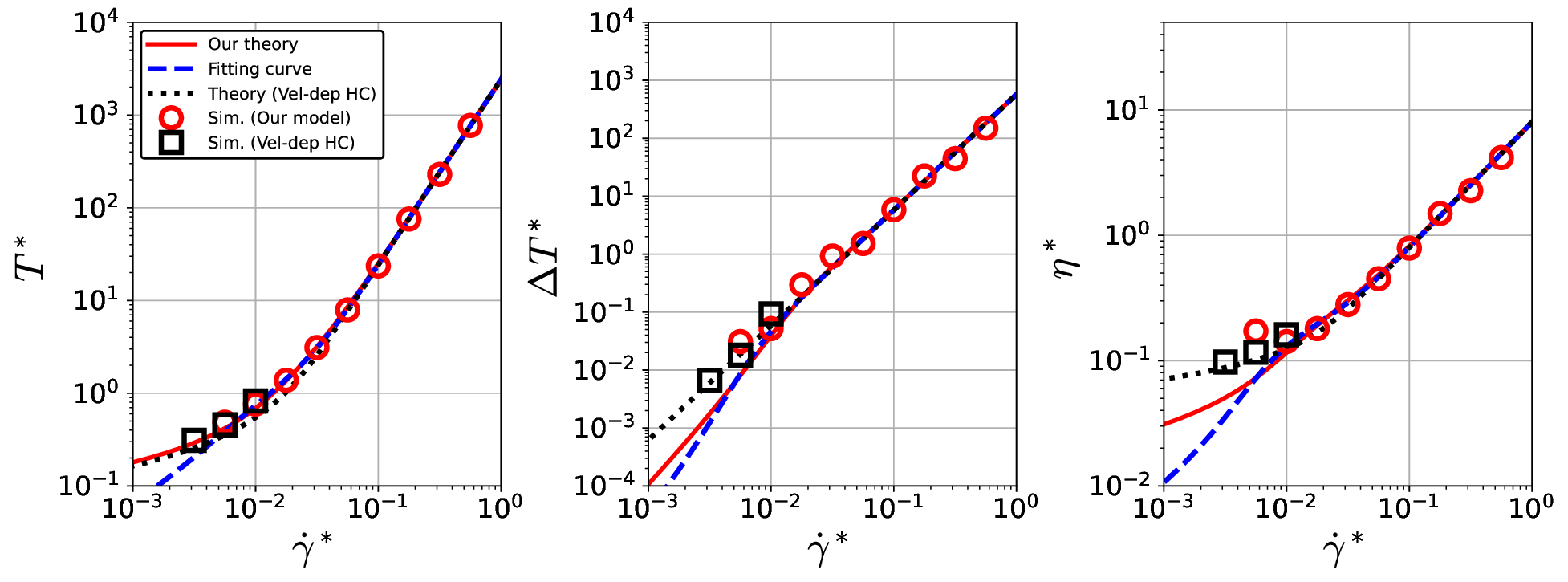}
    \caption{Shear rate dependence of (left) the temperature, (middle) the anisotropic temperature, and (right) the viscosity for $\alpha=4$ and $e=0.9$.
    Here, the solid, dashed, and dotted lines represent the theoretical curve, the fitting curves, and the theoretical curves for velocity-dependent restitution coefficient hard-core model, respectively.
    The markers represent the simulation results from the direct simulation Monte Carlo.}
    \label{fig:rheology}
\end{figure*}

\section{Steady rheology}\label{sec:rheology}
In this section, we present the results for the steady-state rheology.
From Eqs.~\eqref{eq:evol_eqs}, the steady state values are given by
\begin{subequations}\label{eq:steady_values}
\begin{align}
    &\dot\gamma
    = \nu \sqrt{\frac{3}{2}\frac{\zeta}{\nu-\zeta}},\quad
    P_{xy}
    = -\frac{nT}{\nu}
    \sqrt{\frac{3}{2}\zeta(\nu-\zeta)},\\
    &\Delta T 
    = \frac{3\zeta}{\nu}T,\quad
    \eta \equiv -\frac{P_{xy}}{\dot\gamma}
    = nT\frac{\nu-\zeta}{\nu^2}.
\end{align}
\end{subequations}
Before proceeding, it is important to clarify the physical condition under which the steady uniform shear flow (USF) is realized.
The steady state is determined from the energy balance condition, namely, that the viscous heating term $-2/(3n)\dot{\gamma}P_{xy}$ is exactly balanced by the collisional contribution $-\Lambda_{\alpha\alpha}/(3n)$.
Unlike the conventional inelastic hard-sphere model with constant restitution coefficient, the collisional moment $\Lambda_{\alpha\alpha}$ in the present model is not a simple function of temperature, since it reflects the velocity dependence of the effective restitution coefficient.
Therefore, for given values of the core restitution coefficient (at $r=d$) and the imposed shear rate $\dot{\gamma}$, the steady temperature $T$ is determined implicitly from the energy balance equation.
In this sense, $\dot{\gamma}$ and the restitution parameter are externally controlled and can be regarded as independent, while the reduced shear rate $\dot{\gamma}^*$ follows from the resulting steady-state temperature.

Figure~\ref{fig:rheology} shows the dependence of the temperature, temperature anisotropy, and viscosity on the shear rate.
Here, we have employed the dimensionless quantities as 
\begin{equation}
    T^*\equiv \frac{T}{\varepsilon},\ 
    \Delta T^*\equiv \frac{\Delta T}{\varepsilon},\ 
    \dot\gamma^*\equiv \dot\gamma d\sqrt{\frac{m}{\varepsilon}},\ 
    \eta^*\equiv \eta \frac{d^2}{\sqrt{m\varepsilon}}.
\end{equation}
In the high shear regime, the flow curves converge to the following Bagnold scaling~\cite{Santos04, Kobayashi25}:
\begin{subequations}\label{eq:Bagnold}
\begin{align}
    T^{(\mathrm{HC})}
    &= \frac{5\left(2+e\right)}
    {12\pi\left(1-e\right)\left(1+e\right)^2
    \left(3-e\right)^2}
    \frac{m}{n^2 d^4}\dot\gamma^2,\\
    \Delta T^{(\mathrm{HC})}
    &= \frac{25\left(2+e\right)}
    {12\pi\left(1+e\right)^2\left(3-e\right)^3}
    \frac{m}{n^2 d^4}\dot\gamma^2
    ,\\
    \eta^{(\mathrm{HC})}
    &= \frac{5\sqrt{15}\left(2+e\right)^{3/2}}
    {36\pi \left(1-e\right)^{1/2} \left(1+e\right)^2\left(3-e\right)^3}
    \frac{m}{nd^4}\dot\gamma,
\end{align}
\end{subequations}
However, for intermediate and low-shear regime, deviations emerge due to the suppression of low-velocity collisions by the repulsive potential.
The onset of these deviations occurs at $T\simeq \varepsilon$, corresponding to the kinetic energy becoming comparable to the potential barrier.

We also present the direct simulation Monte Carlo (DSMC)~\cite{Bird, Poschel} results in Fig.~\ref{fig:rheology}.
(For details of the DSMC method, see, e.g., Ref.~\cite{Montanero05}.)
These results show good agreement over a wide range of shear rates, demonstrating the broad applicability of our theoretical framework.

\begin{figure}[htbp]
    \centering
    \includegraphics[width=\linewidth]{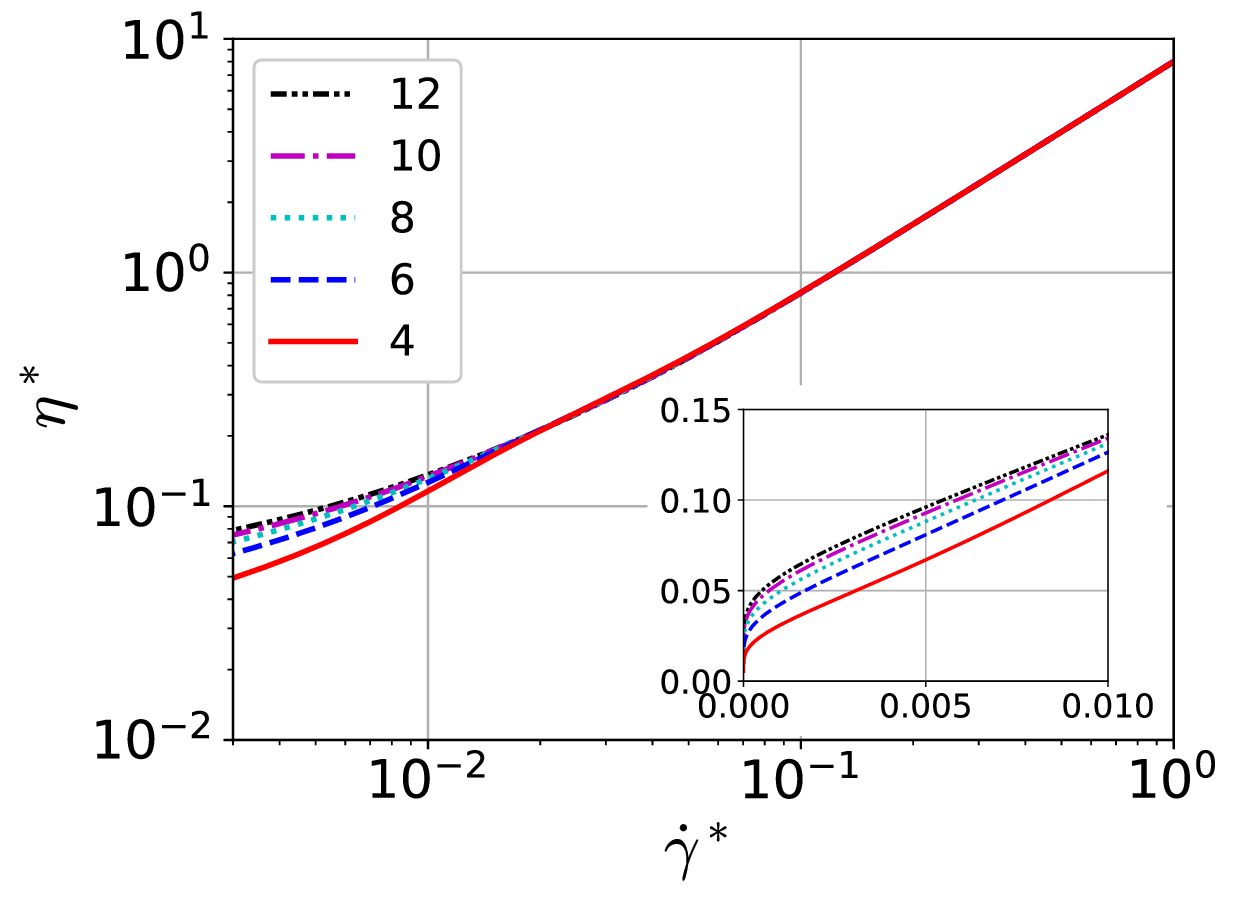}
    \caption{Shear rate dependence of the the viscosity for $\alpha=4$, $6$, $8$, $10$, $12$ when we fix $e=0.9$.
    The inset shows a magnified view of the low-shear region.}
    \label{fig:rheology_alpha}
\end{figure}
Figure~\ref{fig:rheology_alpha} shows the shear rate dependence of the reduced viscosity for fixed $e=0.9$ and several values of the inverse-power-law exponent $\alpha=4$, $6$, $8$, $10$, and $12$.
As in Fig.~\ref{fig:rheology}, the results become almost independent of $\alpha$ in the regime $\dot\gamma^* \gtrsim 0.2$, corresponding to $T \gtrsim \varepsilon$. 
In this high-shear regime, the kinetic energy 
dominates over the repulsive potential barrier, and the system approaches 
the hard-core limit. 
Consequently, the influence of the exponent $\alpha$ becomes negligible and the flow curves asymptotically approach the Bagnold 
scaling.
In contrast, a clear dependence on $\alpha$ emerges in the low-shear regime $\dot\gamma^* \lesssim 0.2$. 
In this region, the viscosity increases with increasing $\alpha$. 
This behavior reflects the stronger suppression of low-velocity collisions as the inverse-power-law interaction becomes 
steeper, which enhances the effective resistance to shear.

\subsection{Introduction of a fitting model}
The evaluation of Eq.~\eqref{eq:Omega_def} requires a double integration for each temperature.
Therefore, when Eq.~\eqref{eq:Omega_def} must be evaluated repeatedly, for instance in iterative calculations of transport coefficients, the computational cost becomes significant.

To overcome this difficulty, we introduce simple fitting forms for $\Omega_{\alpha,n}^{(i)}$ that preserve the correct asymptotic behaviors in both the low- and high-temperature limits:
\begin{subequations}\label{eq:Omega_fitting}
\begin{align}
    \Omega^{(1)\mathrm{fit}}_{\alpha,5}
    &= \Omega^{(1)\mathrm{HC}}_{\infty,n}\frac{1}{2}
    \left\{1+\tanh \left[a_1\log \left(\frac{T}{\varepsilon}\right)-b_1\right]\right\},\\
    \Omega^{(1)\mathrm{fit}}_{\alpha,7}
    &= \Omega^{(1)\mathrm{HC}}_{\infty,n}\frac{1}{2}
    \left\{1+\tanh \left[a_2\log \left(\frac{T}{\varepsilon}\right)-b_2\right]\right\},\\
    \Omega^{(2)\mathrm{fit}}_{\alpha,7}
    &= \Omega^{(2)\mathrm{HC}}_{\infty,7}
    \left\{1+\log\left[1+b_3\left(\frac{T}{\varepsilon}\right)^{-a_3}\right]\right\}.
\end{align}
\end{subequations}

The hyperbolic tangent form for $\Omega^{(1)}_{\alpha,n}$ is chosen so as to reproduce the vanishing behavior in the low-temperature regime and the hard-sphere limit $\Omega^{(1)\mathrm{HC}}_{\infty,n}$ in the high-temperature regime, while ensuring a smooth crossover between the two limits.
Similarly, the functional form adopted for $\Omega^{(2)}_{\alpha,7}$ correctly interpolates between the temperature-dependent behavior at low temperatures and the constant asymptotic value at high temperatures.

\begin{figure}[htbp]
    \centering
    \includegraphics[width=\linewidth]{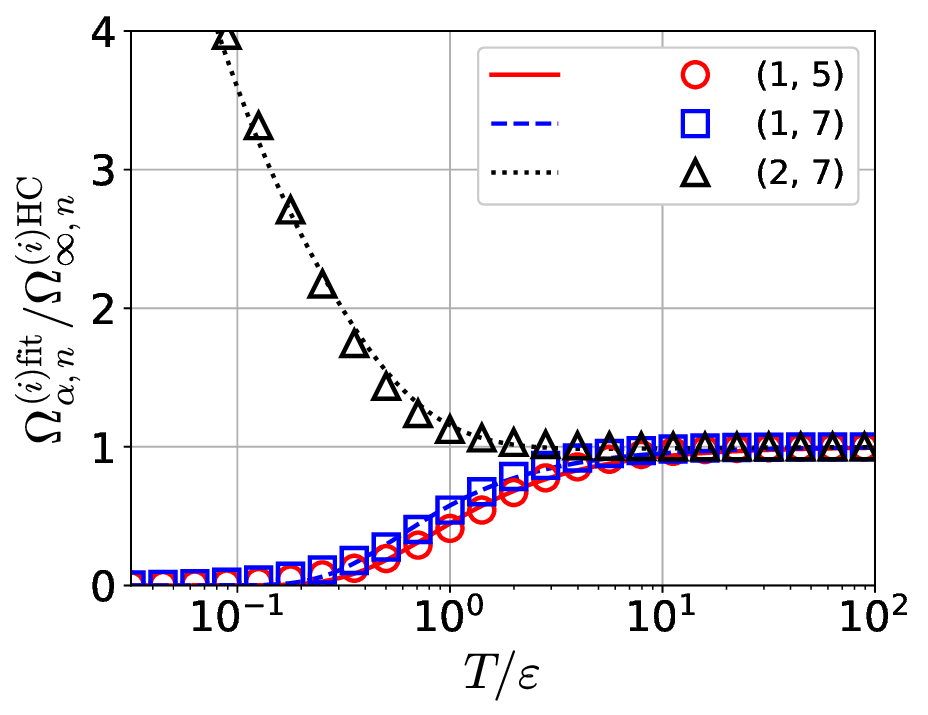}
    \caption{Comparison between $\Omega^{(i)}_{\alpha,n}$ (lines) and the fitting functions $\Omega^{(i)\mathrm{fit}}_{\alpha,n}$ (markers) for $e=0.9$.}
    \label{fig:Omega_fitting}
\end{figure}

The fitting parameters $a_i$ and $b_i$ ($i=1,2,3$) are determined by the least-squares method, yielding
\begin{align}
    a_1&=0.773,\quad b_1=0.232,\quad
    a_2=0.817,\nonumber\\
    b_2&=-0.107,\quad
    a_3=2.04,\quad b_3=0.134.
\end{align}
The fitted curves reproduce the numerical integration results with high accuracy over the entire temperature range considered as shown in Fig.~\ref{fig:Omega_fitting}.

\begin{figure}[htbp]
    \centering
    \includegraphics[width=\linewidth]{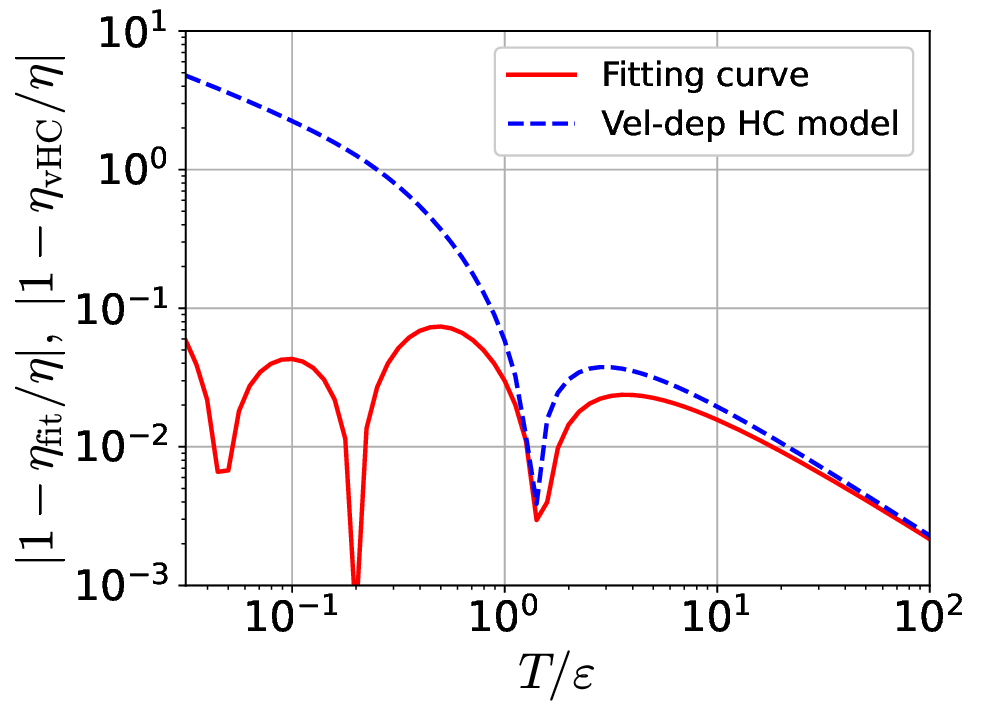}
    \caption{Relative errors of the viscosity obtained from the fitted values of $\Omega^{(i)\mathrm{fit}}_{\alpha,n}$ (solid line) and the velocity-dependent hard-core model (dashed line), measured with respect to the viscosity predicted by our full theory for $e=0.9$.}
    \label{fig:eta_comp}
\end{figure}

The results obtained by substituting Eq.~\eqref{eq:Omega_fitting} into Eq.~\eqref{eq:steady_values} are shown in Fig.~\ref{fig:eta_comp}.
It is found that for $\dot{\gamma}^* \gtrsim 10^{-2.5}$, the fitting model quantitatively reproduces the original results.

\subsection{Comparison with velocity dependent restitution coefficient hard-core model}
In this subsection, let us discuss the difference between our theory and another model.
Some papers~\cite{Poschel03,Takada17,Takada22} introduced the following velocity dependent restitution coefficient hard-core model to mimic charged granular particles:
\begin{equation}
    \mathcal{E}(v_n)=1-(1-e)\Theta(v_n-v_0),
    \label{eq:v_dep_e}
\end{equation}
where $v_0$ is the characteristic speed with respect to the magnitude of charges of particles.
Here, we also introduce the characteristic temperature $T_0$ defined by the characteristic velocity $v_0$ as $T_0\equiv (1/2)mv_0^2$~\cite{Takada17}.
Once we adopt the velocity dependent restitution coefficient \eqref{eq:v_dep_e}, the quantities become
\begin{subequations}\label{eq:Omega_dis1}
\begin{align}
    \Omega_{\mathrm{dis},5}^{(1)}
    &= 2\left(1-e^2\right)
    \left(1+x\right)
    \mathrm{e}^{-x},
    \label{eq:Omega_dis1a}\\
    \Omega_{\mathrm{dis},7}^{(1)}
    &= 4\left(1-e^2\right)
    \left(3+3x+x^2\right)
    \mathrm{e}^{-x},
    \label{eq:Omega_dis1b}\\
    \Omega_{\mathrm{dis},7}^{(2)}
    &= 16
    -4\left(1-e\right)\left(3+e\right)
    \left(1+x\right)
    \mathrm{e}^{-x},
\end{align}
\end{subequations}
with $x\equiv T_0/(2T)$~\cite{Takada17,Takada22}.
Then, we can determine the shear rate, the anisotropic temperature and the viscosity as functions of the temperature analytically.

Figure \ref{fig:rheology} also shows the shear rate dependencies of the temperature, anisotropic temperature, and viscosity determined from Eqs.~\eqref{eq:Omega_dis1}.
It is obvious that these flow curves well reproduce those of our model.

Quantitatively, the relative deviation between the present model and DSMC results in the viscosity remains within $\approx 10\,\%$ for $0.01<\dot\gamma^*<1$.
This level of agreement indicates that the proposed kinetic formulation successfully reproduces the rheology of charged granular gases without invoking ad hoc velocity-dependent restitution parameters.

\section{Velocity Distribution Function}
\begin{figure}[htbp]
    \centering
    \includegraphics[width=\linewidth]{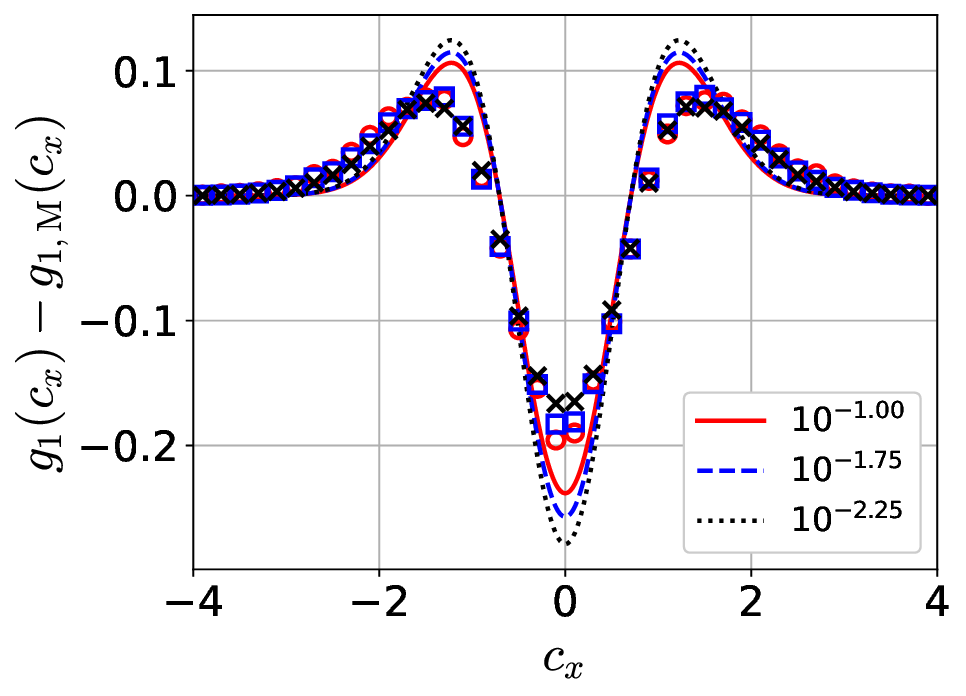}
    \caption{Deviation of the dimensionless velocity distribution function from the dimensionless Maxwellian for shear rates $\dot\gamma^*=10^{-1.00}$ (solid), $10^{-1.75}$ (dashed), and $10^{-2.25}$ (dotted).
    Symbols denote DSMC results.}
    \label{fig:VDF_comp}
\end{figure}

We investigate how the velocity distribution changes in each regime of the flow curve.
For the analysis, we introduce the dimensionless velocity $\bm{c}$ and the dimensionless distribution function $g(\bm{c})$ as
\begin{equation}
    \bm{c}
    \equiv \sqrt{\frac{m}{2T}}\bm{V},\qquad
    g(\bm{c})
    \equiv \frac{v_\mathrm{T}^3}{n}f(\bm{V}),
\end{equation}
where we omit the time variable $t$ since we focus solely on the steady-state distribution.
Although the velocity distribution is originally a function of three variables $(c_x,c_y,c_z)$, for simplicity we consider the dimensionless one-dimensional distribution function obtained by integrating over $c_y$ and $c_z$ \cite{Takada23}:
\begin{align}
    g_1(c_x)
    &\equiv \int_{-\infty}^\infty \mathrm{d}c_y
    \int_{-\infty}^\infty \mathrm{d}c_z
    g(\bm{c}) \nonumber\\
    &= \frac{1}{\sqrt{\pi}}\mathrm{e}^{-c_x^2}
    \left[\frac{1}{2}+\frac{1}{3}\frac{\Delta T}{T}
    +\left(1-\frac{2}{3}\frac{\Delta T}{T}\right)c_x^2\right]\nonumber\\
    &= \frac{1}{\sqrt{\pi}}\mathrm{e}^{-c_x^2}
    \left[\frac{1}{2}+\frac{\zeta}{\nu}
    +\left(1-\frac{2\zeta}{\nu}\right)c_x^2\right].
\end{align}
Figure~\ref{fig:VDF_comp} shows the resulting profiles.

These results indicate that the velocity distribution under shear deviates from the Maxwellian form primarily through the suppression of low-velocity particles and a slight enhancement around the thermal velocity $v\sim v_\mathrm{T}$. 
The deviation remains almost independent of the shear rate, suggesting that the non-Newtonian rheology originates mainly from the anisotropy of the second moments rather than from strong non-Gaussianity of the velocity distribution itself. 
This behavior contrasts with that of neutral granular gases, where velocity-dependent restitution often induces pronounced high-energy tails.

\section{Conclusion}\label{sec:conclusion}
In summary, we have developed a kinetic-theory framework for the rheology of dilute charged granular gases by incorporating both hard-core and inverse-power-law interactions.
The model reproduces Bagnold scaling in the high-shear regime and predicts significant deviations in the intermediate and low-shear regime where the repulsive interaction suppresses inelastic collision frequency.
Theoretical predictions are in quantitative agreement with DSMC, validating the approach.
Future work will extend this formulation to denser regimes, include screening and charge redistribution, and explore the coupling between shear flow and charge dynamics.

\section*{Acknowledgements}
This work is partially supported by the Grant-in-Aid of MEXT for Scientific Research (Grant No.~\href{https://kaken.nii.ac.jp/en/grant/KAKENHI-PROJECT-24K06974/}{JP24K06974}, No.~\href{https://kaken.nii.ac.jp/en/grant/KAKENHI-PROJECT-24K07193/}{JP24K07193}, and No.~\href{https://kaken.nii.ac.jp/en/grant/KAKENHI-PROJECT-25K01063/}{JP25K01063}).

\appendix
\section{Explicit expression of the collision angle $\theta$ for some values of $\alpha$}\label{sec:theta}
In this Appendix, we present explicit expressions of the collision angle $\theta$ for some representative values of the exponent $\alpha$.
For the case where the potential is given purely by an inverse power law, it is well known (see, e.g., Refs.~\cite{Goldstein, Takada24}) for which values of $\alpha$ the integrals can be carried out explicitly.
Although in the present model a rigid-core cutoff is introduced, the algebraic structure of the integrals remains unchanged, except that the lower limit of integration is modified.
Below we illustrate the cases $\alpha=4$ and $6$.

\subsection{Case for $\alpha=4$}
We first consider the case $\alpha=4$.
Let us consider the following equation:
\begin{equation}
    1-x^2-\left(\frac{x}{\tilde{b}}\right)^4=0,
    \label{eq:x0_eq_4}
\end{equation}
which admits two solutions,
\begin{equation}
    x_1^2\equiv \frac{\tilde{b}^2\left(-\tilde{b}^2+ \sqrt{\tilde{b}^4+4}\right)}{2},\quad
    x_2^2\equiv \frac{\tilde{b}^2\left(\tilde{b}^2+ \sqrt{\tilde{b}^4+4}\right)}{2},
\end{equation}
with $x_1\le x_2$.
Using these, Eq.~\eqref{eq:x0_eq_4} can be factorized as
\begin{equation}
    1-x^2-\left(\frac{x}{\tilde{b}}\right)^4
    = \frac{1}{\tilde{b}^4}(x_1^2-x^2)(x_2^2+x^2).
\end{equation}
Accordingly, the collision angle $\theta_4$ becomes
\begin{equation}
    \theta_4
    = \tilde{b}^2\int_0^{x_0}
    \frac{\mathrm{d}x}{\sqrt{(x_1^2-x^2)(x_2^2+x^2)}}.
\end{equation}
To evaluate this integral, we introduce the substitution
\begin{equation}
    x^2 = x_1^2(1-y^2),
\end{equation}
which gives
\begin{equation}
    \mathrm{d}x
    = -x_1 \frac{y}{\sqrt{1-y^2}}\mathrm{d}y.
\end{equation}
This transforms the integral into the standard elliptic form
\begin{align}
    \theta_4
    &= \frac{\tilde{b}}{\left(\tilde{b}^4+4\right)^{1/4}}\int_{y_0}^1
    \frac{1}{\sqrt{(1-y^2)(1-\nu y^2)}}\mathrm{d}y,
\end{align}
where
\begin{equation}
    \nu \equiv \frac{1}{2}\left(1-\frac{\tilde{b}^2}{\sqrt{\tilde{b}^2+4}}\right).
\end{equation}

When the trajectory does not reach the rigid core ($r=d$), we have $x_0=x_1$, which implies $y_0=0$.
In this case, the collision angle $\theta_4$ is expressed as the complete elliptic integral of the first kind:
\begin{align}
    \theta_4
    &= \left(\frac{\tilde{b}^4}{\tilde{b}^4+4}\right)^{1/4}
    \int_0^1
    \frac{1}{\sqrt{(1-y^2)(1-\nu y^2)}}\mathrm{d}y\nonumber\\
    &= \left(\frac{\tilde{b}^4}{\tilde{b}^4+4}\right)^{1/4}
    K(\nu),
\end{align}
where $K(\nu)$ denotes the complete elliptic integral of the first kind \cite{Abramowith}.

On the other hand, when the trajectory does reach the core, the upper limit is $x_0=b/d$.
The corresponding $y_0$ is given by
\begin{equation}
    y_0 = \sqrt{1-\frac{x_0^2}{x_1^2}}
    = \sqrt{1-\left(\dfrac{4\varepsilon}{mv^2}\right)^{1/2}\frac{2}{-\tilde{b}^2+ \sqrt{\tilde{b}^4+4}}}.
\end{equation}
The collision angle $\theta_4$ then takes the form
\begin{align}
    \theta_4
    &= \left(\frac{\tilde{b}^4}{\tilde{b}^4+4}\right)^{1/4}
    \int_{\sqrt{1-\frac{x_0^2}{x_1^2}}}^1
    \frac{1}{\sqrt{(1-y^2)(1-\nu y^2)}}\mathrm{d}y\nonumber\\
    &= \left(\frac{\tilde{b}^4}{\tilde{b}^4+4}\right)^{1/4}
    \left[K(\nu) - F\left(\sin^{-1}y_0,\nu\right)\right],
\end{align}
where $F(\phi, \nu)$ denotes the incomplete elliptic integral of the first kind \cite{Abramowith}.

\subsection{Case for $\alpha=6$}
We next consider the case $\alpha=6$.
In this case, the turning point is determined by the cubic equation in $x^2$:
\begin{equation}
    1-x^2-\left(\frac{x}{\tilde{b}}\right)^6=0.
    \label{eq:x0_eq_6}
\end{equation}
This equation can in principle be solved by the cubic formula or by applying Vieta's formula.
For convenience, we define
\begin{equation}
    \mathcal{F}(X)\equiv 1-X-\frac{X^3}{\tilde{b}^6},
\end{equation}
which is a strictly decreasing function.
Since
\begin{equation}
    \mathcal{F}(0)=1> 0,\quad
    \mathcal{F}(1)=-\frac{1}{\tilde{b}^6}<0,
\end{equation}
there exists a unique real solution in the interval $0<X<1$.
We denote this solution by $x_1^2$.

To bring the integral into standard form, we introduce the change of variables
\begin{equation}
    y\equiv \frac{1}{x^2}-\frac{1}{3}.
\end{equation}
In terms of $y$, the collision angle can be expressed as
\begin{align}
    \theta_6
    = \int_{y_0}^\infty 
    \frac{\mathrm{d}y}{\sqrt{4y^3-\dfrac{4}{3}y-\dfrac{8}{27}-\dfrac{4}{\tilde{b}^6}}}
    = \wp^{-1}(y_0; g_2, g_3),
\end{align}
where $y_0$ is the value corresponding to $x_0$, and $\wp^{-1}(x; g_2, g_3)$ denotes the inverse of the Weierstrass elliptic $\wp$-function with invariants $g_2$ and $g_3$ \cite{Abramowith}.

The value of $x_0$ depends on whether the trajectory hits the hard-core as shown in the main text.
By substituting the appropriate $x_0$, one can therefore evaluate the collision angle $\theta_6$ explicitly in terms of the Weierstrass elliptic function.


\end{document}